\documentclass{article}
\usepackage{amsmath}
\usepackage{amssymb}
\usepackage{epsfig}
\usepackage{a4wide}
\usepackage{multicol}
\usepackage[symbol*]{footmisc}
\begin{document}
%------------------------------------------------------------------
\newcommand{\be}{\begin{equation}}
\newcommand{\ee}{\end{equation}}
\newcommand{\ba}{\begin{eqnarray}}
\newcommand{\ea}{\end{eqnarray}}
%------------------------------------------------------------------
\begin{flushright}\rm July 2015
		   \end{flushright}
\begin{center}
{\Large\bf
Insights into $Q^2 \bar{Q}^2$ states from an effective perspective}\\[1cm]

{\bf Ll. Ametller$^{a}$ and P. Talavera$^b$}\\[0.5cm]
{$^a$ Departament de F\'\i sica i Enginyeria Nuclear,
Universitat Polit\`ecnica de Catalunya, \\
Jordi Girona 5, E-08034 Barcelona,  Spain\\
$^b$ Universitat Internacional de Catalunya,\\
Immaculada 22, E-08017 Barcelona, Spain} 

\begin{abstract}
We discuss the two photon coupling of the lightest scalar meson on the basis of an extension of $\chi$PT. Using 
low-energy data on the pion form-factor and the $\gamma\gamma\to \pi^+\pi^-(\pi^0\pi^0)$ cross-sections as inputs, we find
$\Gamma(\sigma\to\gamma\gamma) \cong 0.126~\rm{keV} $. The smallness of the result and the relative weight between its components, ${\Gamma_{\gamma\gamma\to S_1}\over \Gamma{ \gamma\gamma\to\pi\pi\to S_1}} \le 1$, suggests that the scalar $0^{++}$ meson is mainly a $Q^2\bar{Q}^2$ state.  
\end{abstract}

 % 13.20.Eb, 11.30.Rd, 12.39.Fe, 12.38.Aw}
\end{center}

{\bf Motivation:}~The plethora of scalar mesons in QCD has a long and puzzling history. 
%There are some rather well established scalar mesons: with $I=1$ we find the $a_0(980)$, with $I=1/2$ the $K^*_0(1410)$ and with $I=0$
%the family of the $f_0$ mesons located at $980$, $1379$ and $1500$ MeV. 
Probably, due to its elusiveness, the most interesting state is the isoscalar $I=0$, $\sigma(600)$ \cite{TH1}. It is  well known as a broad enhancement in very low-energy s-wave meson-meson scattering.
The quark or gluon content of the $\sigma(600)$ is not fully understood and the proliferation of models with seemingly different conclusions is disturbing \cite{SR,SR2}. At the same time its underlying structure is a corner stone in understanding the realization of the mechanism for chiral symmetry breaking.

In this paper we find indications of the $Q^2\bar{Q}^2$ content for the $\sigma$ state and estimate it by studying the processes $\gamma\gamma \to \pi\pi$ and the pion vector form-factor. The tetraquark structure of the lightest scalar was proposed  long time ago owing to a possible strong diquark correlation \cite{jaffe}. Our working framework runs in parallel to that in  \cite{ellis2}
with the only difference that we interpret their Lagrangian in an effective perspective by providing a counting power to the singlet field \cite{Soto:2011ap}. Our main result is based on the comparison of two terms: the first one, already studied  in \cite{Ametller:2014vba}, is given by the rescattering effects contribution to the $\gamma\gamma\to \pi\pi\to S_1$ decay and the second by the direct $\gamma\gamma\to S_1$ coupling.
At the fundamental level the two-photon coupling for a generic $S_1$ scalar meson is given by
\begin{equation}
\label{gcoup}
{\cal L}= -\frac{e^2}{ 4 F} c_{1\gamma}  S_1 F_{\mu\nu}F^{\mu\nu}\,.
\end{equation}
There are many ways to couple the scalar singlet to the vacuum. If one considers that the spontaneously breaking of scale invariance is mediated via the trace of the energy-momentum tensor 
the coupling $c_{1\gamma}$ is related to the scalar decay constant via
the relation \cite{ellischanowitz}
\begin{equation}
\label{Rs}
 -\frac{e^2}{ 4 F} c_{1\gamma} F_{S_1} = \frac{\alpha}{6\pi}\frac{\sigma(e^+e^-\to hadrons)}
{ \sigma(e^+e^-\to \mu^+\mu^-)}\quad  \rm{and}\quad \langle 0\vert\theta^\mu_\mu\vert S_1\rangle=-M_{S_1}^2 F_{S_1}\,,
\end{equation}
being $\theta^\mu_\mu$ the trace of the energy momentum tensor
\begin{equation}
\label{trace}
\theta_\mu^\mu:= \theta_g+\theta_q=\frac{1}{4} \beta(\alpha_s) G_{\mu\nu}^a G^{\mu\nu a}+\sum_i m_i\bar{\psi}_i\ (1+\gamma_m(\alpha_s)) \psi_i\,.
\end{equation}
Instead, if we consider that the scalar meson is a S-wave bound state of diquark-antidiquark pair the corresponding interpolating field can be constructed as 
\begin{equation}
j_\sigma=\epsilon_{abc}\epsilon_{dec} (u^T_a C \gamma_5 d_b)(\overline{u}_d  \gamma_5 C \overline{d}^T_e)\,,
\end{equation}
where latin indices denote color and $C$ stands for the charge conjugation matrix. In the above expression the diquark is taken to be a spin zero color antitriplet and flavor antitriplet  \cite{jaffewil}.
Then the coupling to the vacuum is given by \cite{pascual}
\begin{equation}
 \langle 0\vert j_\sigma \vert S_1\rangle=-\sqrt{2} M_{S_1}^4 F_{S_1}\,.
\end{equation}

\hspace{0.5cm}

{\bf Setting the scheme:} Let us first recall the main ingredients of the theoretical set-up. We shall consider an effective approach to QCD with two flavors in the isospin limit. The smallness of the values of the light-quark masses and the external momenta set a 
perturbative scheme out of  the chiral symmetry  limit.  We count the pion and scalar field as 
${\cal O}(p^0)$, derivatives, vector and axial-vector external currents as ${\cal O}(p)$ and the scalar, pseudo-scalar external currents and scalar mass as ${\cal O}(p^2)$. With this counting 
the leading order Lagrangian reduces to that presented in \cite{lanik} 
\begin{eqnarray}
\label{newsigmapion}
{\mathcal{L}}_{2}[0^{++}] =&&
\left( \frac{F^2}{4} +F c_{1d} S_1+ c_{2d} S_1^2+\cdots \right) \langle u_\mu^\dagger  u^\mu \rangle\nonumber \\
+&& \left( \frac{F^2}{4} +F c_{1m} S_1+ c_{2m} S_1^2+\cdots\right) \left( \langle \chi_+\rangle -\langle \chi^\dagger + \chi \rangle\right)\,.
\end{eqnarray}
The pseudo-scalar field is parametrized by the unitary matrix
$
u(x)^2=U(x) = e^{i\sqrt{2} \sum_j \sigma_j\phi_j(x)/F}\,.
$
Here $F$ is the pion decay constant ($F\simeq93$\, MeV), the $\phi_i$'s are fields for the pseudo-scalar Goldstone mesons and $\sigma_i$ are the Pauli matrices.
The basic building blocks are defined as
\begin{eqnarray}
&&\chi= 2 B_0\, (s+i p)\,,\quad\nonumber \\&&
u_\mu=i u^\dagger D_\mu U u^\dagger = - i u D_\mu U^\dagger u=u_\mu^\dagger\,,\nonumber \\&&
 \chi_+=u^\dagger \chi u^\dagger+u\chi^\dagger u\,.
\end{eqnarray}

Next-to-leading order corrections, ${\cal O}(p^4)$, come either through one-loop graphs or by higher order operators. In particular, the terms relevant to our study are explicitly\footnote{To avoid confusion between the low-energy constants in $\chi$PT and S$\chi$PT the former are denoted by $l_i$ while the latter by $\ell_i$. The $\ell_i$ constants become  $l_i$ in the absence of $S_1$.  As is customary in $\chi$PT, the finite and scale independent terms of $\ell_i$ ($l_i$) are denoted by $\bar{\ell}_i$ ($\bar l_i$).} 
\begin{equation}
\label{sphotons}
{\mathcal{L}}_{4}[0^{++}] 
 =  \sum_{i=5}^6 \ell_i P_i + Z_1 \mathring{M}_\sigma^2\langle\chi^\dagger U + \chi U^\dagger\rangle+Z_2 \mathring{M}_\sigma^2 \langle D_\mu U D^\mu U^\dagger \rangle -\frac{e^2}{4 F }    c_{1 \gamma} S_1F_{\mu\nu} F^{\mu\nu} \,,
\end{equation}
where $\mathring{M}_\sigma$ stands for the singlet mass in the chiral limit and
\begin{equation}
\label{polis}
P_5=-\frac{1}{2} \langle f_-^{\mu\nu}f_{-\mu\nu}\rangle\,,\quad P_6=\frac{i}{4} \langle f_+^{\mu\nu}[u_\mu,u_\nu]\rangle\,,
\end{equation}
with
$
f_\pm^{\mu\nu}=uF_L^{\mu\nu}u^\dagger \pm u^\dagger F_R^{\mu\nu} u\,.
$
The field strength tensors $F_{L,R}^{\mu\nu}$ are related to the non-abelian external fields \cite{Gasser:1983yg}. One salient feature of the field theory approach presented above is that it allows to separate between the direct and rescattering $\gamma\gamma$ couplings in a crystal clear fashion. 
The reason being that the direct process always involves the gauge invariant operator in (\ref{sphotons}), $S_1F_{\mu\nu} F^{\mu\nu}$.  
Such separation is not always feasible using dispersion relations.

\hspace{0.5cm}

{\bf Charged pion-pair production:} The amplitude for the process $\gamma(q,\lambda)\gamma(q^\prime,\lambda^\prime)\to \pi^+(p)\pi^-(p^\prime)$, is given by ${\cal A}(\lambda,\lambda^\prime) = e^2\epsilon^\mu(q,\lambda)\epsilon^{\prime\nu}(q^\prime,\lambda^\prime) V_{\mu\nu}^C$, where the $V_{\mu\nu}^C$ tensor can be decomposed into four Lorentz invariant tensor structures although by gauge invariance only two of them have non-vanishing contribution to the 
cross-section
\begin{eqnarray}
\label{structures}
&&V_{\mu\nu}^C= A^C(s,t,u) T_{1\mu\nu} + B^C(s,t,u) T_{2\mu\nu}\,,\nonumber\\&&
T_{1\mu\nu}=\frac{s}{2} g_{\mu\nu}-q_{\nu} q_{\mu}^\prime\,,\quad
T_{2\mu\nu}= 2 s\Delta_\mu\Delta_\nu-\nu^2g_{\mu\nu}-2\nu(q_{\nu}\Delta_\mu-q_{\mu}^\prime\Delta_\nu)\,,
\end{eqnarray}
with $s=(q+q^\prime)^2$, $t=(q-p)^2$, $u=(q-p^\prime)^2$ and $ \Delta_\mu:=(p-p^\prime)_\mu$.
The amplitudes $A^C(s,t,u) $ and $B^C(s,t,u) $ are analytic functions of the Mandelstand variables and are symmetric under crossing 
$\{t,u\}\leftrightarrow\{u,t\}$.
Comparison with the experimental data will be at the level of cross-section. The differential cross-section for unpolarized photons can be casted in terms of the helicity amplitudes $H^C_{+\pm}$ corresponding to helicity changes  $\lambda=0,2$ respectively
\begin{equation}
\frac{d\sigma}{d \Omega} = \frac{\alpha^2 s}{32} \beta(s) H^C(s,t)\,,\quad H^C(s,t)=\vert H^C_{++}\vert^2+\vert H^C_{+-}\vert^2\,.
\end{equation}
In terms of the amplitudes $A^C$ and $B^C$ they read
\begin{eqnarray}
\label{Hs}
&&H^C_{++}=A^C+2(4M_\pi^2-s)B^C\,,\quad
H^C_{+-}=\frac{8(M_\pi^4-t u)}{s} B^C\,.
\end{eqnarray}

At ${\cal O}(p^2)$ there is no scalar contribution and the 
amplitude coincides with that of scalar electrodynamics.
At ${\cal O}(p^4)$ we have found remarkably many more diagrams than in $\chi$PT.  Their evaluation is rather straightforward and the contributions can be conveniently cast in terms of two tensorial structures as
\begin{eqnarray}
{\cal A}^{(4)}= e^2 A(s,t,u) ( s \epsilon\cdot \epsilon^\prime-2 q\cdot \epsilon^\prime\, q^\prime\cdot \epsilon)+e^2 B(s,t,u) \left( \epsilon\cdot \epsilon^\prime-\frac{ \epsilon\cdot p\,\epsilon^\prime\cdot p^\prime}{q\cdot p}-\frac{ \epsilon\cdot p^\prime\, \epsilon^\prime\cdot p}{q\cdot p^\prime}\right)\,,
\end{eqnarray}
which  are related to those in (\ref{structures}) by
\begin{eqnarray}
\label{AB}
&&A^C(s,t,u)= 2A(s,t,u)+\frac{B(s,t,u)}{2}\left(\frac{1}{M_\pi^2-t}+\frac{1}{M_\pi^2-u}\right)\,,\nonumber\\
&&B^C(s,t,u)=\frac{B(s,t,u)}{4 s}\left(\frac{1}{M_\pi^2-t}+\frac{1}{M_\pi^2-u}\right)\,.
\end{eqnarray}
We have performed several checks on our full expressions:
$i)$ In the evaluation we have not fixed neither an specific gauge nor a system of reference and hence we are able to check explicitly gauge invariance in the results.
$ii)$ All non-local divergences cancel when adding the full set 
of diagrams together with wave function renormalization. 
$iii)$ The polynomial
divergences also cancel against the counter-terms determined
in  \cite{Ametller:2014vba}
$
2 \gamma_5=\gamma_6= \frac{1}{3}(4 c_{1d}^2-1)\,.
$
$iv)$ Once we shift the bare pion mass to the renormalized physical one the amplitude turns to be independent of $Z_1$ and $ Z_2$. 
In view of these stringent checks, we trust our calculations
of the matrix-elements.

At this ${\cal O}(p^4)$ order each of the above amplitudes can be split  as
\begin{eqnarray}
\label{desc}
A(s,t,u) &=&  \left[  \frac{1}{F^2} \left\{ 2( 2{\ell}_5-{ \ell}_6) +
\overline{ G}(s)\right\}\right]
+ \frac{c_{1d}^2}{F^2}   \tilde A_s(s,t,u)  +\frac{c_{1d} c_{1\gamma}}{F^2}
  \tilde A_{\gamma\gamma}(s,t,u)\,,\nonumber \\
B(s,t,u) &=&   \left[ 2 \right]+ \frac{c_{1d}^2}{F^2} \tilde B_s(s,t,u)\,.
\end{eqnarray}
The explicit expressions for the $ \tilde A_s(s,t,u)$,  $ \tilde B_s(s,t,u)$ and $\tilde A_{\gamma\gamma}(s,t,u)$ terms are gathered in the Appendix.
The contributions in squared brackets correspond to $\chi$PT \cite{Bijnens:1987dc}. Notice that corrections to $B(s,t,u)$ at ${\cal O}(p^4)$ are absent in $\chi$PT and only show up at higher orders \cite{Burgi:1996qi}. 
This confirms,  as previously remarked in \cite{Ametller:2014vba}, that the value for observables in S$\chi$PT at ${\cal O}(p^4)$ lie within the ${\cal O}(p^4)$ and ${\cal O}(p^6)$ results in $\chi$PT. 

\begin{figure}[h]
\begin{center}
\epsfig{file=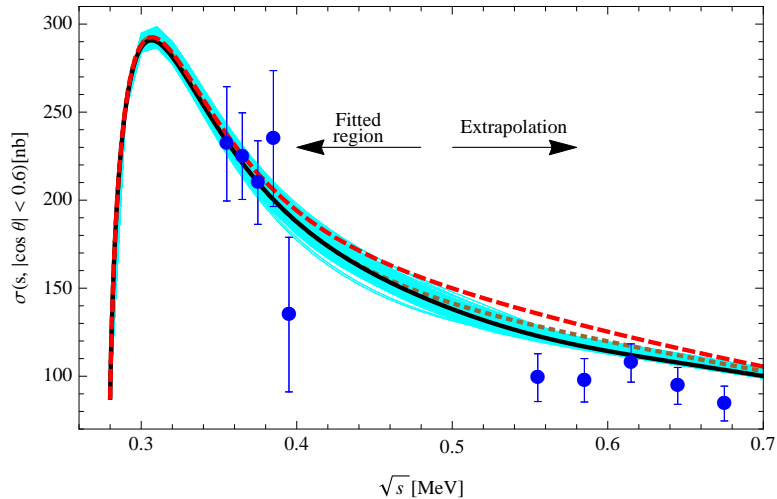,width=0.7\textwidth}
\caption{$\gamma \gamma \to \pi^+\pi^-$ cross-section vs. the center-of-mass energy. Full line corresponds to 
the input values obtained in our fit as described in the main text (\ref{point1}), dashed line corresponds to $\chi$PT at ${\cal O}(p^4)$. Dotted line corresponds to the results for the singlet  obtained in \cite{ccl} where in addition to those we fitted the value of $2{\ell}_5-\ell_6=0.0030$.
We used data values below $0.5$ MeV for our fit, while the rest of the curve is just extrapolated.} 
\label{csection}\end{center}
\end{figure} 

\hspace{0.5cm}

{\bf Results:}  
%The inclusion of electromagnetic terms will hopefully render closer the theoretil prediction to the experimental data.
To extract the value of the $\gamma\gamma S_1$ coupling constant we have simultaneously fitted the experimental central values of the data for 
the processes
$\pi\to\pi\gamma\,, \gamma\gamma\to\pi^0\pi^0$ and $\gamma \gamma\to \pi^+\pi^-$. For the latter 
we only take into account the data points in \cite{mark} near the two-pion production, $\sqrt{s}\approx 0.45$ MeV, this removes to a large extent 
the $K\overline{K}$ effects. The data treatment of the former two experiments is described at lengthly in \cite{Ametller:2014vba}. In all the procedure the only new free parameter, besides $c_{1\gamma}$, at play with respect to those entering in \cite{Ametller:2014vba} is the low-energy constant ${ \ell}_5$.  
We have generated a sufficient refined lattice for the set of constants, $5 \times 10^6$ points, in the
hyperplane defined by $\{c_{1d},M_\sigma,\Gamma^\prime,\ell_\Delta,\bar{\ell}_6, c_{1\gamma}\}$\footnote{Notice that  $\ell_\Delta$ is finite and scale independent. It can be expressed in terms of $\bar{\ell}_i$ quantities as $\ell_\Delta=-{1\over 96\pi^2}\bar{\ell}_\Delta =-{1\over 96\pi^2} (\bar{\ell}_5-\bar{\ell}_6)$.} 
with a 
priori flat distribution and computed their corresponding $\chi^2$ augmented function.
Notice that we have treated all the coupling constants entering in the processes at the same footing, {\it i.e.} without imposing a priori any hierarchy, and nevertheless the output is
consistent with the assumed counting power, $\vert c_{1\gamma}\vert \ll \vert c_{1d}\vert$.
%In doing so
%we assume that the ball park of the contribution is given in the first step.
The main result of this fit is given by
 \begin{eqnarray}
 \label{point1}
& &c_{1d}=0.26_{-0.07}^{+0.10}\,,\quad \ell_\Delta=2\ell_5-\ell_6=0.0026_{-0.0004}^{+0.0015}\,, \quad
\bar{\ell}_6= 19.8_{-2.9}^{+10}\,,\quad \nonumber \\
& &M_\sigma= 553_{-114}^{+46}~\rm{MeV}\,, \quad \Gamma^\prime= 295_{-157}^{+229}~\rm{MeV}\,,
\quad c_{1\gamma}= -0.012_{-0.010}^{+0.016}\,,
 \end{eqnarray}
and is depicted in fig. \ref{csection} as the full curve for the $\gamma\gamma\to\pi^+\pi^-$ cross-section. Fits to 
$\gamma\gamma\to\pi^0\pi^0$ and $\pi\to\pi\gamma$, not shown here, are similar to those obtained in \cite{Ametller:2014vba}.
The total $\chi^2_{\it d.o.f}$ for the joint fit of all the three processes is $\frac{180.4}{69}$. For comparison, the same fit
but using $\chi$PT at ${\cal O}(p^4)$ gives $\chi^2_{\it d.o.f}=\frac{361.4}{65}$.
Notice that the finding concerning $c_{1\gamma}$ matches the short distance arguments that suggest a small two photon coupling \cite{ellis1}. 
Errors in (\ref{point1}) correspond to the 1$\sigma$ deviations.  It is worth emphasizing that the narrow thickness of the band in fig. \ref{csection} suggests that this experiment is not suitable to pin down the scalar mass and/or width. 
This statement is more evident if we compare our outputs for the singlet mass and width, (\ref{point1}), with those obtained in \cite{ccl}, $M_\sigma= 420$ MeV and $\Gamma^\prime =286$ MeV \footnote{We have rewritten the outputs of \cite{ccl}, where mass and width were defined as 
$s_\sigma=\left(M_\sigma - i {\Gamma^\prime\over 2}\right)^2\,,$ in our convention $s_\sigma=M_\sigma^2 - i {\Gamma^\prime M_\sigma}$\,.}. The latter, depicted as the  dotted line in fig. \ref{csection}, lies within the $1\sigma$ deviation from the central value of (\ref{point1}). 
It is also worth emphasizing  that a tiny variation in the fit, for instance including or not the data point at $\sqrt{s}=0.395$ MeV, changes the preferable $\{M_\sigma, \Gamma^\prime\}$ set point that minimizes the data. 
 
The value of the combination of low-energy constants must be compared to those standard estimates obtained  in \cite{Burgi:1996qi}  $ 2l_5-l_6=0.0028$ and  \cite{Gasser:2006qa} $ 2l_5-l_6=0.0031$. 
Or to that
extracted 
independently from the $\pi^+\to e^+\mu_e\gamma$ decay via 
the axial--vector-to-vector form factor ratio $\frac{h_A}{ h_V}$ 
$  2 l_5-l_6=0.0031$ \cite{BT}.
The difference between those results and the corresponding one in (\ref{point1}) gives an understanding of the effect of the singlet field in this combination of low-energy constants. 
As was expected from the beginning the contribution of the scalar singlet is mild in this process because it is  mainly saturated by Vectors and Axials. 

\begin{figure}[h]
\begin{center}
\epsfig{file=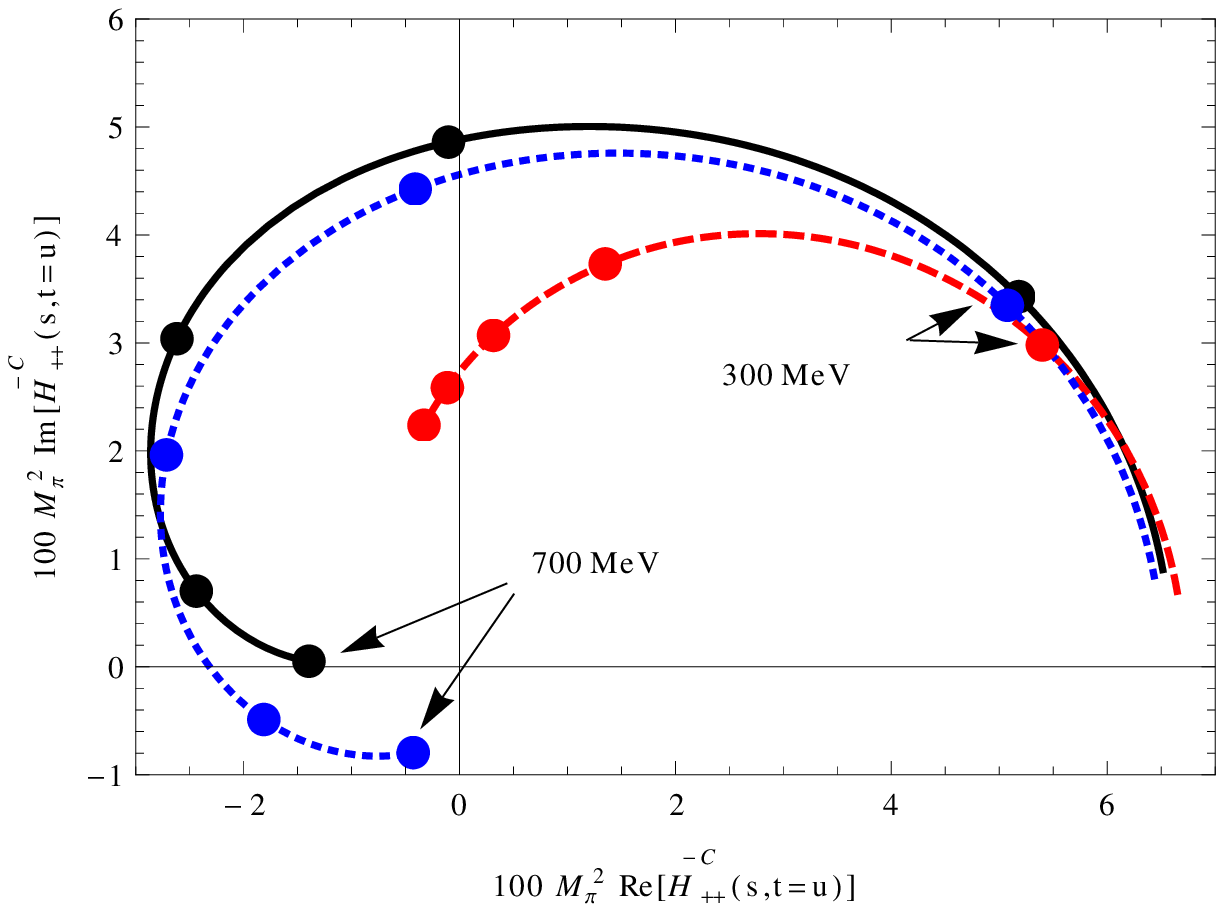,width=0.5\textwidth}
\epsfig{file=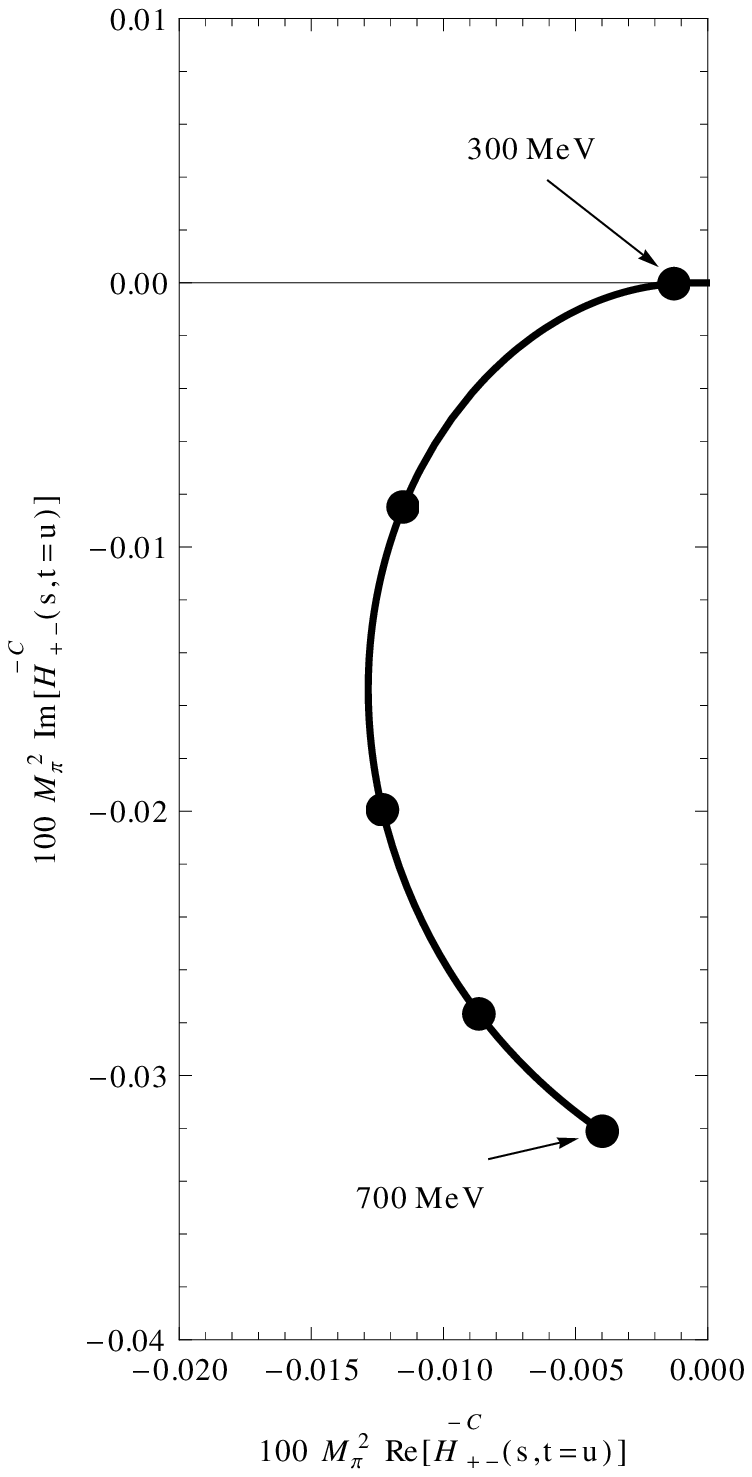,width=0.32\textwidth}
\caption{Imaginary vs. real (Argand plot)  parts of the Born subtracted ${\cal O}(p^4)$ helicity amplitudes, $\overline{H}_{++}^C(s,t=u)$ and  $\overline{H}_{+-}^C(s,t=u)$, as a function of the center-of-mass energy. The solid line is obtained using the central 
values in (\ref{point1}). The dotted curve is as above but setting to zero the electromagnetic coupling. Finally the dashed line
corresponds to the $\chi$PT case. The dots signal the center-of-mass 
energy of the two-pion system in $100$~MeV steps.  Notice that $\overline{H}_{+-}^C$ does not receive any contribution from $\chi$PT at ${\cal O}(p^4)$. Also the dashed line is indistinguishable from the solid line in this latter case.} 
\label{ammpp}
\end{center}
\end{figure} 

In fig. \ref{ammpp} we plot the imaginary vs. the real  parts
of the helicity amplitudes at $t=u$ once the Born contribution is subtracted 
\begin{equation}
\overline{H}_{++}^C:=H_{++}^C-H_{B++}^C\,,\quad
\overline{H}_{+-}^C:=H_{+-}^C-H_{B+-}^C\,.
\label{amsub}
\end{equation}
It is evident that the electromagnetic correction is small and that at large energies there is a relatively large enhancement, with respect to the $\chi$PT, due to the inclusion of the scalar particle. 

As pass by we have also evaluated the dipole polarizabilities of the charged pion. This is obtained via the Compton scattering process $\gamma\pi^+\to\gamma\pi^+$ which is related to the pion-pair production by crossing symmetry $s \leftrightarrow t$. 
Expanding (\ref{amsub}) at the Compton threshold and using the input (\ref{point1}),
we obtained
\begin{eqnarray}
\label{polas}
&&(\alpha_1-\beta_1)_{\pi^+}\cong \left( 4.0^{+2.3}_{-1.3}\,,~[6.0]\,,~\{ 5.7\}\right) \times 10^{-4}~\hbox{fm}^3\,, 
\nonumber \\
&&(\alpha_1+\beta_1)_{\pi^+} \cong \left( 0.012^{+0.011}_{-0.006}\,,~[0]\,,~\{ 0.16\}\right)  \times 10^{-4}~\hbox{fm}^3\,,
\end{eqnarray}
where the numbers in square (curly) brackets stand for the standard $\chi$PT values at 
${\cal O}(p^4) ({\cal O}(p^6))$ respectively.
%The numerical difference with respect to $\chi$PT values is beyond the experimental accuracy.
As in the $\chi$PT case, it seems very hard to reconciliate  the sharp discrepancy of 
(\ref{polas}) with the most recent experimental result based on the radiative pion photo-production, $\gamma p\to \gamma\pi^+ n$, 
$
(\alpha_1-\beta_1)_{\pi^+}^{\rm exp}= (11.6\pm 1.5_{\rm{stat.}}\pm 3.0_{\rm{syst.}}\pm 0.5_{\rm{mod.}})\times{10^{-4}~\hbox{fm}^3}  \cite{exp}\,.
$

\hspace{0.5cm}

{\bf Revisiting $\Gamma(S_1\to\gamma\gamma)$:}~
We are now in a position of finding the decay width of the scalar singlet to two photons. This was partially treated in \cite{Ametller:2014vba} with the proviso that its direct coupling to photons was suppressed and the bulk of the contribution comes from the radiative process. Relaxing the above assumption and taking the $\gamma\gamma S_1$ term into account we obtain 
 \begin{equation} 
 \label{swidh}
\Gamma(S_1\to\gamma\gamma) =  \frac{\alpha^2 \pi }{4 F^2} M_\sigma^3  \vert  c_{1\gamma}-8  c_{1d} \frac{M_\sigma^2-2M_\pi^2}{M_\sigma^2}
\overline{G} (M_\sigma^2)\vert^2\cong 0.126_{-0.044}^{+0.349}~\rm{keV}\,.
\end{equation} 

%\begin{multicols}{2}

Notice that in the previous expression both terms, Born and radiative corrections, 
are of the same effective counting power.
Analytically (\ref{swidh}) agrees with the Born approximation of \cite{ellis2} once we set
$c_{1d}=0$. 
It is worth emphasizing the $M_\sigma^3$ dependence in the above expression. This makes specially relevant the definition of the mass for a particle which width and mass are comparable. Had we used the convention in \cite{ccl} our prediction for the central value of the scalar mass would have been approximately a $3\%$ larger, or, equivalently, the sigma radiative width would increase a factor $\simeq 1.1$. This can be accounted for as a source of  systematic error. 

Owing to the smallness of (\ref{swidh}) in comparison with the characteristic width of  a conventional 
$Q\bar{Q}$ resonance, for instance $\Gamma(f\to \gamma\gamma)\approx 5-6\,\rm{keV}$ \cite{t5}, we can conclude at the light of  (\ref{trace})  that $S_1$ is mainly non-$Q\bar{Q}$. A comparison with other results that can be found in the literature is collected in table 1.  One salient point is that ours is roughly  a decade lower than the results obtained through dispersive calculations.
% and agrees, ball park,
% with other analysis based on an effective approach but including a larger range of energies $\approx 1.5 \text{GeV}$ \cite{t2} or other phenomenological models based in Vector Meson Dominance \cite{t2a}.

\begin{table}[h]
\label{sca}% is used to refer this table in the text
\centering % used for centering table
\begin{tabular}{c c c} % centered columns (2 columns)
&\text{Reference}&$\Gamma(S_1\to\gamma\gamma)\,\,\left[\text{keV}\right]$\\
\hline%\hline %inserts double horizontal lines
&\text{This work}&$0.126_{-0.044}^{+0.349}$\\
& \cite{SR2}  &$2.8$\\
%&\cite{t10}& $0.22$\\
%&\cite{t14}&$0.27$ \\
%&\cite{t2}  & $0.47\pm 0.66$\\
%&\cite{t2a}  & $0.024\pm0.023;\, 0.38\pm0.09$\\
&\cite{t1} & $1.68\pm0.15$\\
&\cite{t3} & $2.08\pm 0.20$\\
&\cite{t4}& $1.4\sim 3.2$\\
&\cite{puff}& $3.08$\\
&\cite{t6}& $2.08$\\
%&\cite{t11}& $0.2\sim 0.3$\\ 
&\cite{t12}&$1.7\pm 0.4$\\
&\cite{t13}&$1.2\pm 0.4$\\
&\cite{t15}&$4$ \\
&\cite{t16}\,\text{model A}&$3.5\pm 0.7$\\
&\cite{t16}\,\text{model B}&$2.4\pm 0.5$ \\
%&\cite{t17}&$4.1\pm 0.3$\\
&\cite{t18}&$\le 1$ \\[0.5ex] % inserts table
%heading
% [1ex] adds vertical space
\hline %inserts single line
\end{tabular}
\caption{Comparison for $\Gamma(S_1\to\gamma\gamma)$ between different models.} % title of Table
\end{table}

Aiming at a further theoretical interpretation we have checked whether 
this deviation w.r.t. the dispersive calculation can be assessed to a strong $\sigma \to K \bar{K}$ coupling \cite{puff}.  We have extended our analytical results to $SU(3)$ and for a first and very crude estimation we took {\sl naively} at face the values given in (\ref{point1}) considering different ratios for the quantity $r_{\sigma K\pi}={g_{\sigma KK}\over g_{\sigma \pi\pi}}$\footnote{ $g_{\sigma \pi\pi}$ stands for the obvious generalization of $c_{1d}$.}. By looking at the results, collected in table 2, we may conservatively expect almost no sensitivity in the singlet decay width due to the presence of the strange quark mass.  
\begin{table}[h]
\label{ratio}% is used to refer this table in the text
\centering % used for centering table
\begin{tabular}{c c c c c c} % centered columns (2 columns)
&$r_{\sigma K\pi}$ &$1$ [\text{universality}]& $0.8$ \cite{SR2} &$0.37$ \cite{t4}& $0.62$ \cite{t6}   \\
\hline%\hline %inserts double horizontal lines
&$\Gamma(S_1\to\gamma\gamma)\,\,\left[\text{keV}\right]$& $0.134$&$0.132$ &$0.128$ &$0.130$\\[0.5ex] % inserts table
%heading
% [1ex] adds vertical space
\hline %inserts single line
\end{tabular}
\caption{Comparison for an $SU(3)$ extension of $\Gamma(S_1\to\gamma\gamma)$ using a naive extrapolation for the low-energy constants.} % title of Table
\end{table}

\begin{figure}[h]
\begin{center}
\epsfig{file=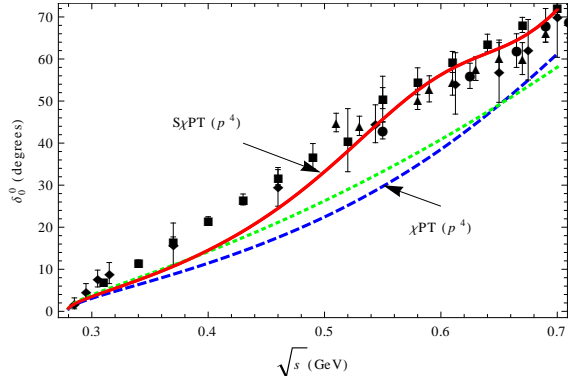,width=0.5\textwidth}
\caption{$\gamma\gamma\to \pi^+\pi^-$, $I=0$ s-wave  phase shift $\delta_0^0$ in the low-energy region. The keys to experimental data are as follows: $ \blacksquare=\cite{datasq}\,, \blacklozenge=\cite{datasqq}\,,\blacktriangle=\cite{datatr}\,, \bullet=\cite{datacr}$.  The blue dashed and red full lines correspond to the  ${\cal O}(p^4)$ result for  $\chi$PT and S$\chi$PT respectively.  
The green dotted line denotes  the $\pi\pi$ elastic scattering $\delta_0^{0(2)}$.
All curves agree at threshold, obeying Watson's theorem. }
\label{delta00}
\end{center}
\end{figure}

In order to cross-check further our full approach we have computed the $\gamma\gamma\to \pi^+\pi^-$ $I=0$ s-wave phase shift $\delta_0^0$ in the threshold region. Those are related to the $\pi\pi$ elastic scattering phase shifts through Watson's theorem. We have proceeded reconstructing the partial waves amplitudes, $T_l^I$, from the neutral and charged $\gamma\gamma\to \pi\pi$ processes and through these the phase shifts.  As is customary we express
the $\pi-\pi$ elastic scattering result  as the expansion in energy \cite{juerg}
\begin{equation}
\delta_l^I= \arctan\left( \text{Re} T_l^I\right) + {\cal O}(E^6)
%=\text{Re} T_l^{I(2)} +\text{Re} T_l^{I(4)}+ {\cal O}(E^6)
=\delta_l^{I(2)} +\delta_l^{I(4)}+ {\cal O}(E^6)\,,\quad 16 M_\pi^2 \ge s\ge 4 M_\pi^2\,.
\end{equation}
At any time we bear in mind that the truncated chiral expansion becomes unreliable above $\sqrt{s}\approx 450$ MeV. 
%In the other side the direct experimental data on $\delta_0^0$ is rather poor below $\sqrt{s}\approx 600$ MeV. 
In fig. \ref{delta00} we have depicted our results for the central values (\ref{point1}) adding for comparison the corresponding $\chi$PT ones. As is evident from the figure we obtain a remarkable improvement w.r.t. the $\chi$PT prediction and the agreement with the most recent data is rather good specially for the energy range 
$0.5\,\text{GeV}\le \sqrt{s}\le 0.7\, \text{GeV}$. We stress that there is no fit to these data and is just a {\sl prediction} or a consistency check. This together with the fact that we reproduce the experimental data for the $\pi-\pi$ scattering lengths, pion polarizabilities and the pion radii \cite{Ametller:2014vba} let us to think that we have obtained a fairly good parameterization of the low-energy region containing the effects of the singlet state.

%\end{multicols}
Once we have settled the consistency of the approach let us come back to the discussion on the quark content. 
The gluonium or four-quark scenarios are the most controversial scenarios to disentangle from our analysis. 
%In fact by simply inspection and comparison with the values quoted in table 6 of \ref{t10} one can conclude that the scalar can be anything exception of a $Q\bar{Q}$ state.  
In order to do so
we look at the relative weight between both terms in (\ref{swidh}). Considering just the direct coupling term one obtains the results in table 3.
\begin{table}[h]
\label{direct}% is used to refer this table in the text
\centering % used for centering table
\vspace{0.25cm}
\begin{small}
\begin{tabular*}{\textwidth}{*{11}{c@{\hspace{1.7mm}}}c}
%\begin{tabular}{c c c c c c c c c c} % centered columns (2 columns)
& &\text{This work}&  \cite{SR2} &\cite{t4}& \cite{puff}\ & \cite{ACH}& \cite{t8}&   \cite{t9}& \cite{t2a}\,{\text{Fit I}}& \cite{t2a}\,{\text{Fit II}}\\
\hline% \hline %inserts double horizontal lines
&$\Gamma_{S_1\to \gamma\gamma}\,\,\left[\text{keV}\right]$& $0.115_{-0.115}^{+0.114}$&$0.13\pm0.05$ &$0.3$&$0.16$ &$0.005$&$0.05-0.1$&$0.9$&  $0.024\pm0.023$&$ 0.38\pm0.09$\\[0.5ex] % inserts table
%heading
% [1ex] adds vertical space
\hline %inserts single line
\end{tabular*}
\end{small}
%\end{tabular}
\caption{Comparison of different results for the direct contribution to the decay width.} % title of Table
\end{table}

Considering  instead  the decay via the rescattering process we get the results in table 4.
\begin{table}[h]
\label{rsc}% is used to refer this table in the text
\centering % used for centering table
\begin{tabular}{c c c c c c } % centered columns (2 columns)
& &\text{This work}&  \cite{SR2} &\cite{puff}\ & \cite{acha}\\
\hline%\hline %inserts double horizontal lines
&$\Gamma_{S_1\to \pi\pi\to\gamma\gamma}\,\,\left[\text{keV}\right]$& $0.194^{+0.282}_{-0.113}$&$2.7\pm 0.4$ &$1.89$ &$2$\\[0.5ex] % inserts table
%heading
% [1ex] adds vertical space
\hline %inserts single line
\end{tabular}
\caption{Comparison of different results for the rescattering contribution to the decay width.} % title of Table
\end{table}
The difference with the $0.11$ keV value found in \cite{Ametller:2014vba} is due to the slightly bigger value of $c_{1d}$.  Thus the initial mismatch with the dispersive calculation can be traced back to the rescattering term. 
In particular the difference can have a twofold origin, see eq. (\ref{swidh}): {\sl i)} The constant $c_{1d}$. Although  this constant is estimated at tree level its value is essentially  upper bounded by the pion vector form-factor and the $I=0$ s-wave phase shift $\delta_0^0$ below the $K\bar K$ threshold. One expects that the value is renormalized and at the scale $\mu$ is enhanced by a factor $\log\left({M_\pi^2\over \mu^2}\right)$. {\sl ii)} The $\overline{G} (M_\sigma^2)$ function, or more generically pion rescattering effects. Notice that due to counting power contributions to $\Gamma_{S_1\to \gamma\gamma}$ start already at ${\cal O}(p^4)$  thus one expects higher order corrections of $\approx 20\%\sim 30\%$. To estimate these we have partially resummed  a subset of higher order diagrams obtaining an increase of $\sim 10\%$ w.r.t. the central value in (\ref{swidh}). Thus, although this result is incomplete seems to indicate that the numerical differences w.r.t. the dispersive results are hard to be asset to higher order corrections. 

On the other side one has to bear in mind that dispersive calculations are not free of uncertainties. Just to mention a few  instances: 
{\sl i)} The results seem to be very sensible to the matrix element parameterization above the $K\bar{K}$ threshold.  
{\sl ii)}  Only the s-wave component is kept at low-energy. 
{\sl iii)}  The result seems to be very sensitive to the actual value of the analogous of $c_{1d}$, i.e. $g_{\sigma\pi\pi}(s)$\footnote{Not to be confused with the energy independent generalization of $c_{1d}$ used in Table 2.}.
For instance the differences
 between the value \cite{t4,t6}, given in table 1 and those found in \cite{t11}\,, $0.2\sim 0.3\,\text{keV}$\,, are just due to the value of this coupling.
{\sl iv)} The approach, by analytical continuation, evaluates the matrix element  deep in the complex plane. One has to keep in mind that the original embedding \cite{omnes} is valid for point like particles which matrix elements are evaluated near the real axis. 

%Furthermore only the s-wave component is kept at low-energy. Just to stress the delicate balance between different %pieces playing a role in the calculation, if one uses dispersive calculations but using a Breit-Wigner cross-section at %low-energy to fit the data, the result agrees with ours within errors \cite{comment},

Comparing the central value for the direct and rescattering decay  widths we learn that the relative weight between both terms in (\ref{swidh})  is approximately $1:2$ and that their interference is partially destructive. The relative smallness of the direct coupling in front of the radiative term, mediated via pion loops, can be interpreted as an indication of a dominant $Q^2\bar{Q}^2$ component in the nature of the scalar singlet. However this conclusion has to be taken cautiously as we have checked that for an increasing singlet mass the scenario can be reversed.  
Obviously all the above reflections are in the absence of mixing which can obscure this simple picture. In fact, 
this is neither strange nor new as similar conclusions are supported by QCD sum-rules  \cite{SR}, lattice QCD calculations \cite{lat} and large N$_c$ scaling arguments \cite{jr}. The novelty of our approach resides in that this finding is encoded in the low-energy regime and an effective approach suffices to capture it.

%This requires further study

%This factor leads us to two conclusions. First, the XXXX is in general dominant over the XXXX.
%The properties of XXX therefore generally determine the 

\hspace{0.5cm}

{\bf Conclusions:} We have found an  estimate to the scalar to two photons
decay width using low-energy data. 
The fact that the preferred point is attained for $\vert c_{1\gamma}\vert  <  1$ but not vanishing signals the presence of the canonical anomaly \cite{ellis1}.
Making use of the central values of (\ref{point1}) together with (\ref{Rs}), and $R=\frac{5}{ 3}$ for consistency,  we obtain 
$F_{S_1}\approx -2.3 F_\pi\,,$ to be compared with the $Q\bar{Q}$ and the pure glueball results: 
$
F_{S_1}=-F_\pi
$
and  $F_{S_1}\approx- 5 F_\pi$ respectively \cite{ellis2}. This fact reinforces our conclusions about the tetraquark nature of the scalar meson as derived from its coupling to two-photons (\ref{swidh}).

Adopting the most optimistic attitude, taking into account higher order resummations, $SU(3)$ extensions and the sensitivity of the results on 
$c_{1d}$ and $M_\sigma$ the central value in (\ref{swidh}) can be pushed up to 
 \begin{equation} 
\Gamma(S_1\to\gamma\gamma)  \approx (0.3 \sim 0.4) ~\rm{keV}\,.
\end{equation} 
This agrees with other effective approaches, studies of the low-energy data using a Breit-Wigner cross-section or
studies of the $\gamma\gamma\to\pi^0\pi^0$ cross-section assuming a scalar dominance \cite{comment}. There is however a mismatch of  a factor $4\sim 5$ w.r.t. the dispersive approaches. 

Concerning the possible sources of difference w.r.t. the dispersive calculations two comments are in order: 
{\sl i)} We have an analytic expression for the rescattering piece at ${\cal O}(p^4)$. We remind that within the effective framework unitarity is only satisfied perturbatively contrary to the dispersive approach where unitary is enforced by construction and use of high-energy data is taken into account. This is also the main reason underlaying the small deviation from the scattering phase shifts data below
$0.5$ GeV as in the standard case \cite{donoghue}. 
{\sl ii)} Concerning the second source, the coupling $c_{1d}$, it has a more controversial status. Being a tree level constant we have found it essentially through  processes in which its role enters at the radiative level. 
Due to the sensitivity in the dispersive approach to the value of $g_{\sigma\pi\pi}(s)$ it would be interesting to have a constraint on $c_{1d}$ in processes where it plays a dominant role even at leading order. 

We stress that only low-energy data were used in our approach.

\vspace{1cm}
\noindent
\noindent{\bf Acknowledgements}

P.T. is partially supported by  FPA2013-46570.

\section*{Appendix}

\begin{figure}[h]
\begin{center}
\epsfig{file=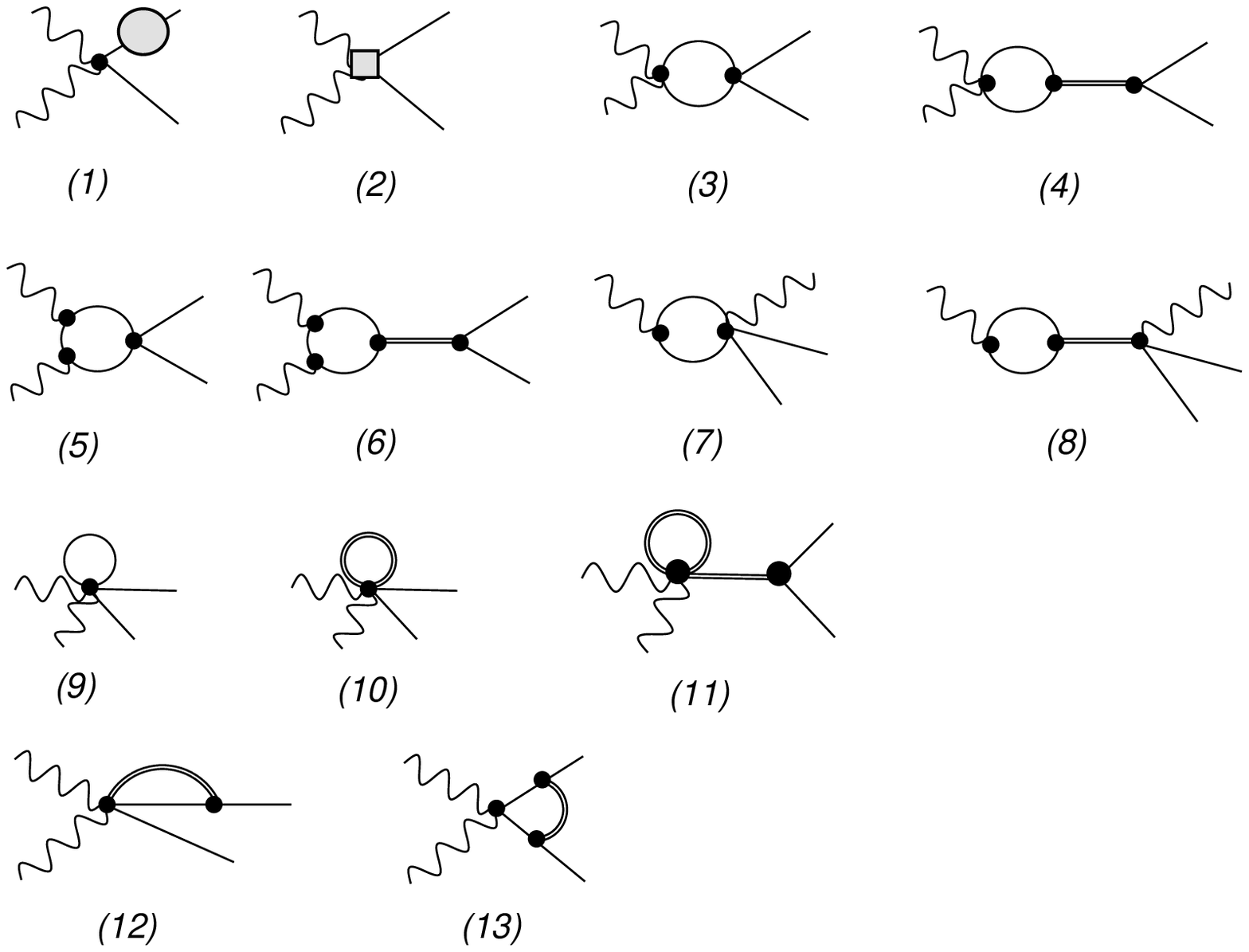,width=0.55\textwidth}
\epsfig{file=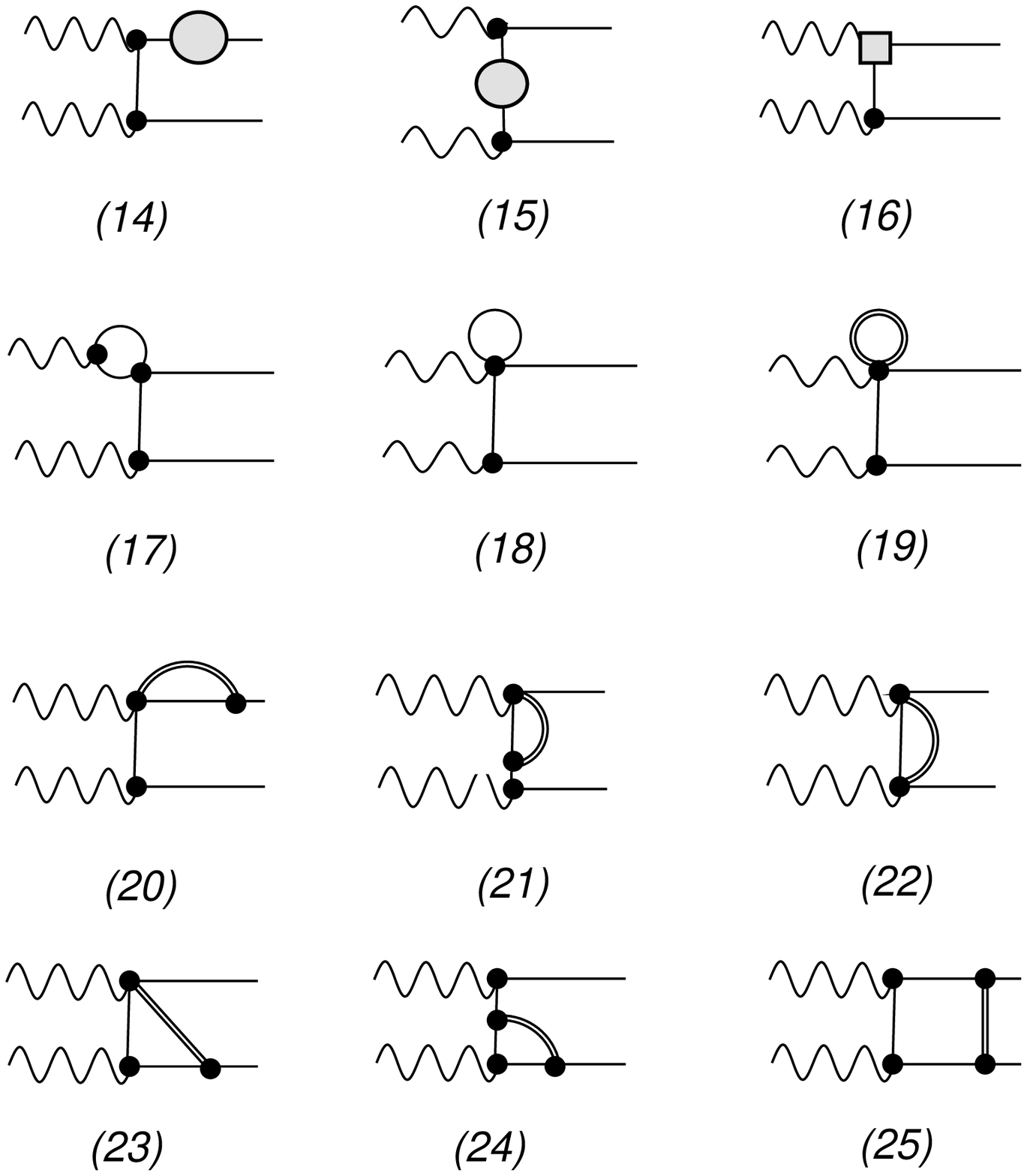,width=0.4\textwidth}
\caption{Feyman diagrams for the process $\gamma\gamma\to \pi^+\pi^-$. The keys to the states are as follows: wavy lines$=$photons, full lines $=$ pions and double full lines $=$ scalar singlet.  Diagrams from (1)-(13) denote the s-channel and they contribute to the $\epsilon\cdot \epsilon^\prime$ component of the amplitude. 
The $t$ ($u$)-channel, diagrams (14)-(25), contribute to the transverse components of the amplitude.}
\label{diagg}
\end{center}
\end{figure}

We have gathered in this appendix all the relevant information concerning the rational functions and integrals appearing in the amplitudes (\ref{desc}) together with the diagrams that describe the process $\gamma\gamma
\to \pi^+\pi^-$, see fig. (\ref{diagg}).
In the calculation we used dimensional regularization in the $\overline{\text{MS}}$ scheme.

The short hand notation of the amplitude can be casted in terms of the finite part of the one-, two-, three- and four-point scalar functions as
\begin{eqnarray}
&&{\cal K}_{0}(s,t,u)=\mu_\pi-\mu_\sigma\,,\quad
{\cal K}_{1}(s,t,u)=\overline{J}_{\pi \sigma }(M_{\pi }^2)\,,\quad
{\cal K}_{2}(s,t,u)=\overline{ J}_{\pi \sigma }(t)\,,\quad\nonumber \\&&
{\cal K}_{3}(s,t,u)= \overline{G}(s) \,,\quad
{\cal K}_{4}(s,t,u)=\overline{C}_0\left(s,M_{\pi }^2,M_{\pi }^2,M_{\pi }^2,M_{\pi }^2,M_{\sigma }^2\right)\,,\quad \nonumber \\&&
{\cal K}_{5}(s,t,u)= \overline{C}_0\left(0,t,M_{\pi }^2,M_{\pi }^2,M_{\pi }^2,M_{\sigma }^2\right)\,, 
  \nonumber \\  &&
 {\cal K}_{6}(s,t,u)= 
    \overline{D}_0\left(0,0,M_{\pi }^2,M_{\pi }^2,s,t,M_{\pi }^2,M_{\pi }^2,M_{\pi }^2,M_{\sigma }^2\right) \,,
   \end{eqnarray} 
where  the  $\overline{C}_0$ and $\overline{D}_0$ functions are the ones introduced in \cite{Passarino:1978jh} and the overline indicates that they incorporate the $\frac{1}{16\pi^2}$ factors, as the $\bar{J}$ and  $\bar{G}$ functions do.  In particular
\begin{equation}
\bar{G}(s) :=  - \left[{1\over 16\pi^2}+ 2 M_{\pi }^2 \overline{C}_0(0,0,s, M_{\pi }^2,M_{\pi }^2,M_{\pi }^2)\right]\,.
\end{equation}  
   
We can split the amplitudes in terms of a polynomial piece and a dispersive one as
\begin{eqnarray}
\label{ammp}
&&\tilde A_s(s,t,u)= \left[P_A(s,t,u)+ \sum_{i=0}^6 U_{Ai}(s,t,u) {\cal K}_i (s,t,u)\right]+\left[t \leftrightarrow u\right]\,, \nonumber  \\
&&\tilde B_s (s,t,u) = \left[P_B(s,t,u)+   \sum_{i=0}^6 U_{Bi}(s,t,u) {\cal K}_i (s,t,u)\right]+\left[t \leftrightarrow u\right]\,,
\end{eqnarray}     
where the $ {\cal K}^\prime$s correspond to the scalar loop functions  and $U^\prime$s are rational functions of the masses, scalar width and Mandelstand variables, with $\nu=t-u$. The terms contributing to the $\tilde A_s(s,t,u)$ contribution are

\begin{eqnarray}
&&
P_A(s,t,u)=\frac{s  \left(M_{\sigma }^2-2
   M_{\pi }^2\right){}^2 \left(2 M_{\sigma }^2-4 M_{\pi
   }^2+s\right)}{16 \pi ^2 \left(M_{\pi }^4-t u\right){}^2}\,,
  \nonumber   \\
  &&  
   U_{A0}(s,t,u)=0\,,
    \nonumber   \\
&&  
 U_{A1}(s,t,u)=   \frac{4    \left(M_{\sigma }^2-2 M_{\pi }^2\right){}^2
   \left(M_{\pi }^4 s+2 M_{\pi }^2 t u-2 M_{\pi }^6+s t u\right)}{ s \left(t-M_{\pi }^2\right) \left(M_{\pi }^2-u\right) \left(t u-M_{\pi }^4\right)}\,,\   
   \nonumber    \\
   &&
U_{A2}(s,t,u)=   -\frac{8   t   \left(M_{\sigma }^2-2 M_{\pi }^2\right){}^2 \left(M_{\pi }^2-u\right) }{ s \left(t-M_{\pi }^2\right) \left(M_{\pi }^4-t u\right)}\,,\nonumber \\
 &&
 \begin{aligned}
U_{A3}(s,t,u)= &\frac{1}{ 8 s \left(M_{\pi }^4-t
   u\right){}^2}\left[16 M_{\pi }^4 s
   \left(10 s M_{\sigma }^2+s^2+2 \nu^2\right)-
   2 M_{\sigma }^2
   \left(s^2-\nu^2\right)^2+s
   \left(s^2-\nu^2\right)^2\right.
    \nonumber    \\
   +& \left.  8
   s^3 M_{\sigma }^4+16 s^2 M_{\sigma }^6
   -64 M_{\pi }^6 s^2-4 M_{\pi }^2
   \left(24 s^2 M_{\sigma }^4+4 s M_{\sigma }^2
   \left(s^2+\nu^2\right)+ s^4-\nu^4\right) \right]  
 \nonumber    \\
   -& 4  \left(s-2 M_{\pi
   }^2\right)^2   {F(s)\over  s}\,, 
    \end{aligned}
    \nonumber    \\
 &&
U_{A4}(s,t,u)= -\frac{2  M_{\pi }^2   \left(M_{\sigma }^2-2 M_{\pi }^2\right){}^2 \left(-2 M_{\pi }^4+t^2+u^2\right)
  }{ \left(M_{\pi }^4-t u\right){}^2}\,, \nonumber    \\
     &&
  U_{A5}(s,t,u)=   -\frac{ 8 M_{\pi }^2   \left(M_{\sigma }^2-2 M_{\pi }^2\right){}^2 \left(M_{\pi }^2-t\right) \left(-M_{\sigma }^2+M_{\pi}^2+t\right) }{ \left(M_{\pi }^4-t u\right){}^2}\,,
   \nonumber    \\
&&
\begin{aligned}
   U_{A6}(s,t,u) =&  -\frac{ 4 M_{\pi }^2   \left(M_{\sigma }^2-2 M_{\pi }^2\right){}^2}{ s \left(M_{\pi }^4-t u\right){}^2} \left[s^2 M_{\sigma }^4+M_{\pi }^4 \left(s^2-12
   t^2\right)-2 s^2 M_{\sigma }^2 \left(M_{\pi }^2+t\right)\right.
\nonumber    \\
   &\left.
   +2 M_{\pi }^2 t^2 (3 s+4 t)
   -2 M_{\pi }^6 (s-4 t)-2 M_{\pi }^8-t^2 \left(s^2+4 s
   t+2 t^2\right)\right] \,,
   \end{aligned}
     \end{eqnarray} 
and the terms contributing to the $\tilde B_s(s,t,u)$ amplitude read
\begin{eqnarray}
&&
\begin{aligned}
P_B(s,t,u)=& \frac{-\left(M_{\sigma
   }^2-2 M_{\pi }^2\right){}^2}{16\pi ^2  M_{\pi }^2 \left(4 M_{\pi
   }^2-M_{\sigma }^2\right) \left(M_{\pi }^4-t
   u\right){}^2}  \left[ 4 M_{\pi }^2 \left(M_{\pi
   }^2 s^3+M_{\pi }^4 \left(-6 s^2+t^2+u^2\right)+8 M_{\pi }^6
   s \right. \right.  \nonumber  \\
&-
 \left.   5 M_{\pi }^8+t u \left(t^2+t u+u^2\right)\right)+2 s
   M_{\sigma }^4 \left(t-M_{\pi }^2\right) \left(M_{\pi
   }^2-u\right)
    \nonumber  \\
   &+ \left. 
   s \left(s-12 M_{\pi }^2\right) M_{\sigma }^2
   \left(t-M_{\pi }^2\right) \left(M_{\pi
   }^2-u\right)\right]\,,
   \end{aligned}
 \nonumber  \\
    &&
 U_{B0}(s,t,u)=  \frac{8  M_{\pi }^2 ( M_{\sigma }^2-2 M_{\pi }^2)^2}{-5 M_{\pi }^2 M_{\sigma }^2+ M_{\sigma }^4+4M_{\pi }^4}\,,
   \nonumber\\
   &&
   \begin{aligned}
 U_{B1}(s,t,u)=&  \frac{4  \left(M_{\sigma }^2-2 M_{\pi }^2\right){}^2}{ M_{\pi }^2 \left(4 M_{\pi }^2-M_{\sigma }^2\right) \left(M_{\pi }^2-t\right) \left(M_{\pi
   }^2-u\right) \left(M_{\pi }^4-t u\right)}
   \left[-M_{\pi }^8 \left(M_{\sigma }^2+s\right)  \right. \nonumber\\
  & \left.  +M_{\pi }^2 t u \left(2 s M_{\sigma }^2-3 t u\right)
   -7 M_{\pi }^4 s t u+t^2 u^2 M_{\sigma
   }^2-2 M_{\pi }^6 t u+5 M_{\pi }^{10}\right] \,,
   \end{aligned}
 \nonumber    \\
   &&
 U_{B2}(s,t,u)=  \frac{8  \left(M_{\sigma }^2-2 M_{\pi }^2\right){}^2 t \left(u-M_{\pi }^2\right)}{  \left(t-M_{\pi }^2\right) \left(t u-M_{\pi }^4\right)}\,,
 \nonumber    \\
  &&
 U_{B3}(s,t,u)=  \frac{ s  \left(M_{\sigma
   }^2-2 M_{\pi }^2\right){}^2 \left(t-M_{\pi }^2\right)
   \left(M_{\pi }^2-u\right) \left(2 M_{\sigma }^2-4 M_{\pi
   }^2+s\right)}{ M_{\pi }^2 \left(M_{\pi }^4-t u\right){}^2}\,,
 \nonumber    \\
   &&
 U_{B4}(s,t,u)=   \frac{  2  \left(M_{\sigma }^2-2 M_{\pi }^2\right){}^2 \left(t-M_{\pi }^2\right) \left(u-M_{\pi }^2\right) \left(-2 M_{\pi
   }^4+t^2+u^2\right) }{  \left(M_{\pi }^4-t
   u\right){}^2}\,,
 \nonumber    \\
 &&  
 U_{B5}(s,t,u)= \frac{ 8   \left(M_{\sigma }^2-2 M_{\pi }^2\right){}^2 \left(t-M_{\pi }^2\right)^2 \left(M_{\pi }^2-u\right)}{ \left(M_{\pi }^4-t u\right){}^2}
 \left(-M_{\sigma }^2+M_{\pi }^2+t\right)\,. \nonumber    \\
   &&
   \begin{aligned}
   U_{B6}(s,t,u)=& -
   \frac{  4  \left(M_{\sigma }^2-2 M_{\pi }^2\right){}^2 \left(t-M_{\pi }^2\right) \left(M_{\pi }^2-u\right)}{  \left(M_{\pi }^4-t u\right){}^2} \left[-2 M_{\pi
   }^2 \left(s M_{\sigma }^2+t^2\right)+s \left(t-M_{\sigma }^2\right){}^2 \right.\nonumber    \\
   &+\left.
   M_{\pi }^4 (s+4 t)-2 M_{\pi }^6\right]\,.
   \end{aligned}
\end{eqnarray}    
Finally, the amplitude proportional to the direct $\sigma\gamma\gamma$ coupling $c_{1\gamma}$ (\ref{desc}) is
\begin{equation}
 \tilde A_{\gamma\gamma}(s,t,u)  = \left({s\over 2}-M_{\pi }^2 \right) F(s)\,.
\end{equation}
Notice that in the above expressions we have used a Breit-Wigner representation to regularize the propagator of the scalar particle
\begin{equation}
F(s)= {1\over s-M_\sigma^2+i M_{\sigma }\Gamma^\prime}\,.
\end{equation}
This can be, at first sight, slightly controversial. The main two arguments to use this parametrization are: {\sl i)} as in all the processes studied in \cite{Ametller:2014vba} in this work the propagator enters in the highest radiative order, thus differences between parametrizations would be reflected at least at ${\cal O}(p^6)$, beyond our scope. This would drastically change in the case of studying $\pi-\pi$ scattering where already the scalar propagator enters at lowest order. {\sl ii)} In this line, we have recovered the results  in  \cite{Ametller:2014vba}, within the $1\sigma$ band,  using a different parameterization \cite{pp}: 
\begin{equation}
F(s) = {1\over s-  M_\sigma^2+i M_{\sigma }\Gamma(s)}\,,\quad \text{ with}\quad \Gamma(s)=
\left( {s-s_0\over M_\sigma^2 - s_0}\right)^{3/2} \Gamma_0\,.
\end{equation}

\end{document}